\documentclass[a4paper,11pt,oneside]{article}

\usepackage[english]{babel}
\usepackage[T1]{fontenc}
\usepackage[utf8]{inputenc}

\usepackage[pdftex,unicode]{hyperref}
\hypersetup{pdftitle=Clustering of Arrivals in Queueing Systems: Autoregressive Conditional Duration Approach}
\hypersetup{pdfauthor=Petra Tomanová and Vladimír Holý}

\usepackage[margin=60pt]{geometry}
\usepackage{amsmath}
\usepackage{amssymb}
\usepackage{bm}

\usepackage[group-minimum-digits=3]{siunitx}

\usepackage{enumitem}

\usepackage{graphicx}
\usepackage{float}
\usepackage{hhline}
\usepackage{multirow}
\usepackage{rotating}
\usepackage{booktabs}

\usepackage[authoryear]{natbib}

\usepackage{xcolor}

\begin{document}

\begin{center}
{\Large \bfseries Clustering of Arrivals in Queueing Systems: \\ Autoregressive Conditional Duration Approach\footnote{Preliminary results were presented in \cite{Tomanova2018}, \cite{Tomanova2019} and \cite{Tomanova2019a}.}}
\end{center}

\begin{center}
{\bfseries Petra Tomanová} \\
University of Economics, Prague \\
Winston Churchill Square 1938/4, 130 67 Prague 3, Czechia \\
\href{mailto:petra.tomanova@vse.cz}{petra.tomanova@vse.cz}
\end{center}

\begin{center}
{\bfseries Vladimír Holý} \\
University of Economics, Prague \\
Winston Churchill Square 1938/4, 130 67 Prague 3, Czechia \\
\href{mailto:vladimir.holy@vse.cz}{vladimir.holy@vse.cz}
\end{center}

\begin{center}
{\itshape \today}
\end{center}

\noindent
\textbf{Abstract:} Arrivals in queueing systems are typically assumed to be independent and exponentially distributed. Our analysis of an online bookshop, however, shows that there is an autocorrelation structure present. First, we adjust the inter-arrival times for diurnal and seasonal patterns. Second, we model adjusted inter-arrival times by the generalized autoregressive score (GAS) model based on the generalized gamma distribution in the spirit of the autoregressive conditional duration (ACD) models. Third, in a simulation study, we investigate the effects of the dynamic arrival model on the number of customers, the busy period, and the response time in queueing systems with single and multiple servers. We find that ignoring the autocorrelation structure leads to significantly underestimated performance measures and consequently suboptimal decisions. The proposed approach serves as a general methodology for the treatment of arrivals clustering in practice.
\\

\\

\\

\section{Introduction}
\label{sec:intro}

In various applications of the operations research, it is undeniable that the characteristics of models evolve over time. The parameters of interest can depend on the time of day and season (see e.g.\ \citealp{KayacCodur2020}) as well as on their past values and other past indicators (see e.g. \citealp{Bruzda2020}). In the paper, we focus on the latter dependency in arrivals to queueing systems from the perspective of the autoregressive conditional duration models with the generalized autoregressive score dynamics.

Many standard queueing systems consider inter-arrival times to be independent due to analytical tractability. Some studies, however, explicitly consider autocorrelation and model arrivals using the \emph{Markovian arrival process (MAP)} (see e.g.\ \citealp{Adan2003, Buchholz2017, ManafzadehDizbin2019}), the \emph{Markov renewal process} (see e.g.\ \citealp{Tin1985, Patuwo1993, Szekli1994}), the \emph{moving average process} (see e.g.\ \citealp{Finch1963, Finch1965, Pearce1967}) or the \emph{discrete autoregressive process} (see e.g.\ \citealp{Hwang2003, Kamoun2006, Miao2013}). \cite{Hwang2003} argue that the time series models with few parameters are more suitable in practice than the MAP models which might be overparameterized. Simulation studies investigating the autocorrelation in arrivals include \cite{Livny1993}, \cite{Resnick1997}, \cite{Altiok2001}, \cite{Nielsen2007} and \cite{Civelek2009}. Overall, these studies show that ignoring the autocorrelation structure, if present, leads to biased performance measures in queueing systems.

The arrival processes are also extensively studied in the financial high-frequency literature. In this field, the duration analysis deals with the modeling of times between successive transactions (trade durations), times until the price reaches a certain level (price durations), and times until a certain volume is traded (volume durations). Typically, the \emph{autoregressive conditional duration (ACD) model} of \cite{Engle1998} is utilized. Its dynamics are analogous to the \emph{generalized autoregressive conditional heteroskedasticity (GARCH) model} of \cite{Bollerslev1986}. In its basic version, the ACD model is based on the exponential distribution but many other distributions are considered in the literature as well. Notably, \cite{Lunde1999a} introduces the generalized gamma distribution to the ACD model. \cite{Bauwens2004} and \cite{Fernandes2005} find that the generalized gamma distribution is more adequate than the exponential, Weibull, and Burr distributions in financial applications. \cite{Hautsch2003} further finds that the four-parameter generalized F distribution reduces to the three-parameter generalized gamma distribution in most cases of financial durations. For a survey of the financial duration analysis, see \cite{Pacurar2008} and \cite{Saranjeet2019}.

A modern approach to time series modeling is the \emph{general autoregressive score (GAS) model} of \cite{Creal2013}, also known as the \emph{dynamic conditional score (DCS) model} by \cite{Harvey2013}. The GAS model is an observation-driven model providing a general framework for modeling of time-varying parameters of any underlying probability distribution. It captures the dynamics of time-varying parameters by the autoregressive term and the score of the conditional density function utilizing the shape of the density function. The theoretical properties of the GAS models together with their estimation by the maximum likelihood method are investigated e.g.\ by \cite{Blasques2014a} and \cite{Blasques2018}. The empirical performance of the GAS models is studied e.g.\ by \cite{Koopman2016} and \cite{Blazsek2020}. So far, there are over 200 papers devoted to numerous models belonging to the GAS family with various applications -- see \cite{GasWeb2020} for a comprehensive list of publications.

The class of the ACD models and the class of the GAS models overlap. In the case of the exponential distribution, the ACD model is equivalent to the GAS model (see \citealp{Creal2013}). For more complex distributions, however, they tend to differ as the ACD models are driven by the lagged observation (or, when rewritten, the difference between the observation and the expected value) while the GAS models are driven by the lagged score. In general, the GAS models are very often superior when compared to alternatives (see e.g.\ \citealp{Blazsek2015, Koopman2016, Chen2019, Gorgi2019, Harvey2019, Blazsek2020}). Concerning GAS models for positive or non-negative values that are suitable for the duration analysis, \cite{Fonseca2018} utilize the Birnbaum--Saunders distribution, \cite{Holy2018} utilize the zero-inflated negative binomial distribution as well as the generalized gamma distribution, and \cite{Harvey2020} utilize the generalized beta distribution as well as the generalized gamma distribution.
 
In the paper, we put together these three cornerstones -- the queueing theory, the duration analysis, and the GAS models -- and demonstrate that they fit together perfectly. The literature already successfully incorporates the GAS models to the duration analysis as discussed above while the perspective from the queueing theory is our novel contribution. We analyze inter-arrival times between orders from an online Czech bookshop. First, we adjust arrivals for diurnal and seasonal patterns using the cubic spline. Next, we find that the adjusted inter-arrival times exhibit strong clustering behavior -- short inter-arrival times are usually followed by short times. To capture this autocorrelation, we utilize the dynamic model based on the generalized gamma distribution with the GAS dynamics in the spirit of the ACD models. We confirm that the proposed specification is quite suitable for the observed data. Finally, we investigate the effects of the proposed arrivals model on queueing systems with single and multiple servers and exponential services. In a simulation study, we show that various performance measures -- the number of customers in the system, the busy period of servers, and the response time -- have higher mean and variance as well as heavier tails for the proposed dynamic arrivals model than for the standard static model. Furthermore, we illustrate how the misspecification of the arrivals model can lead to suboptimal decisions.

The rest of the paper is structured as follows. In Section \ref{sec:arrive}, we present the model based on the generalized gamma distribution with the GAS dynamics for diurnally adjusted inter-arrival times. In Section \ref{sec:emp}, we show that real data of a retail store exhibit an autocorrelation structure that is well captured by our model. In Section \ref{sec:queue}, we investigate the impact of the proposed arrivals model on the performance measures using simulations. We conclude the paper in Section \ref{sec:con}.

\section{Dynamic Model for Arrivals}
\label{sec:arrive}

\subsection{Diurnal and Seasonal Adjustment}
\label{sec:arriveAdj}

Before we utilize the generalized autoregressive score (GAS) model to capture the autoregressive structure of inter-arrival times, we need to deal with diurnal, weekly, and monthly seasonality patterns. To model the non-linear behavior of the diurnal and seasonal patterns and to properly adjust the inter-arrival times, the cubic spline method is utilized. The \emph{cubic spline} is a piecewise cubic polynomial with continuous derivatives up to the order two at each $k$-th fixed point called a knot, $k=1,\dots,K$. \cite{Bruce2017} point out that the cubic spline method is often a superior approach to the polynomial regression since the polynomial regression often leads to undesirable \emph{"wiggliness"} in the regression equation.

To take into account the specifics of raw inter-arrival times $\left\{\tilde{y}_i\right\}_{i=1}^n$, we define the cubic spline with knots at $\left\{\xi_k\right\}_{k=1}^K$ as
\begin{equation}\label{eq:splines}
\log \tilde{y}_i = \beta_0 + \beta_1 b_1(x_i) + \beta_2 b_2(x_i) + \dots + \beta_{K+3} b_{K+3}(x_i) + \gamma t_i + \varepsilon_i, 
\end{equation}
where $\left\{\beta_j\right\}_{j=1}^{K+3}$ and $\gamma$ are parameters to be estimated, $\varepsilon_i$ is disturbance term, $t_i$ is the trend variable, $\left\{b_j\right\}_{j=1}^{K+3}$ are the basis functions, and $x_i$ is a time difference in minutes between the time-stamp of the $i$th observation and the beginning of the week (Monday 00:00) to which the $i$th observation belongs. Thus, $\left\{x_i\right\}_{i=1}^n$ is able to capture both diurnal and intra-week patterns. The basis functions are equal to (i) the variable $x_i$, $b_1(x_i) = x_i$; (ii) its square, $b_2(x_i) = x^2_i$; (iii) its cube, $b_3(x_i) = x^3_i$; and (iv) truncated power functions, $b_{k+3}(x_i) = \max\left\{0, (x_i-\xi_k)^3\right\}, k=1,\dots,K$. The trend variable $t_i$ is linear in time (not linear in observations), $t_1 = 0$ and $t_i = \sum_{j=1}^{i-1} \tilde{y}_j$ for $i = 2,\dots,n$, to take into account the irregularly spaced observations. Moreover, the logarithmic transformation of $\tilde{y}$ ensures the non-negativity of adjusted inter-arrival times. Equidistant intervals are used for knots identification since intervals based on quantiles might lead to a too-small number of knots allocated to off-peak hours.

Regression parameters in \eqref{eq:splines} are estimated by the \emph{weighted least squares (WLS)} method with weights being the inter-arrival times. The WLS naturally compensates for the possibility that during a particular time interval either a small number of long inter-arrival times or a higher number of shorter inter-arrival times is observed, i.e. the number of observed inter-arrival times within a time interval depends on the values of inter-arrival times themselves. Unlike the ordinary least squares, this approach properly weights the inter-arrival times during hours that exhibit a small median but a huge dispersion. Once the parameters are estimated, the diurnally and seasonally adjusted and detrended inter-arrival times $y_i$ are set to exponentiated residuals from regression \eqref{eq:splines}.

\subsection{Generalized Gamma Distribution}
\label{sec:arriveGamma}

Next, we consider the adjusted inter-arrival times $y_i$ to follow the generalized gamma distribution. The \emph{generalized gamma distribution} is a continuous probability distribution for non-negative variables proposed by \cite{Stacy1962}. It is a three-parameter generalization of the two-parameter gamma distribution and contains the exponential distribution and the Weibull distribution as special cases. The distribution has the scale parameter $\alpha$ and the shape parameters $\psi > 0$ and $\varphi > 0$. We use the parametrization allowing for arbitrary values of $\alpha$ which is quite suitable for modeling of its dynamics. The probability density function is
\begin{equation}
\label{eq:gengammaDensity}
f(y | \alpha, \psi, \varphi) = \frac{1}{\Gamma \left( \psi \right) } \frac{\varphi}{e^{\alpha}} \left( \frac{y}{e^{\alpha}} \right)^{\psi \varphi - 1} e^{- \left( \frac{y}{e^{\alpha}} \right)^{\varphi}} \qquad \text{for } y \in (0, \infty),
\end{equation}
where $\Gamma \left( \cdot \right)$ is the gamma function. The expected value and variance is
\begin{equation}
\label{eq:gengammaMoments}
\begin{aligned}
\mathrm{E}[Y] &= e^\alpha \frac{\Gamma \left(\psi + \varphi^{-1} \right)}{\Gamma \left( \psi \right)}, \\
\mathrm{var}[Y] &= e^{2\alpha} \frac{\Gamma \left( \psi + 2 \varphi^{-1} \right)}{\Gamma \left( \psi \right)} - \left( e^{\alpha} \frac{\Gamma \left(\psi + \varphi^{-1} \right)}{\Gamma \left( \psi \right)} \right)^2. \\
\end{aligned}
\end{equation}
The score for the parameter $\alpha$ is
\begin{equation}
\label{eq:gengammaScore}
\nabla_{\alpha} (y, \alpha, \psi, \varphi) = \frac{\partial \log f(y | \alpha, \psi, \varphi)}{\partial \alpha} = \varphi \left( y^{\varphi} e^{-\varphi \alpha} - \psi \right) \qquad \text{for } y \in (0, \infty).
\end{equation}
The Fisher information for the parameter $\alpha$ is
\begin{equation}
\label{eq:gengammaFisher}
\mathcal{I}_{\alpha} (\alpha, \psi, \varphi) = \mathrm{E} \left[ \nabla_{\alpha} (y, \alpha, \psi, \varphi)^2 \middle| \alpha, \psi, \varphi \right] = \psi \varphi^2.
\end{equation}
Note that the Fisher information for $\alpha$ is not dependent on $\alpha$ itself. Special cases of the generalized gamma distribution include the gamma distribution for $\varphi = 1$, the Weibull distribution for $\psi = 1$ and the exponential distribution for $\psi = 1$ and $\varphi = 1$. The generalized gamma distribution itself is contained in a larger family -- the generalized F distribution with four parameters.

\subsection{Generalized Autoregressive Score Dynamics}
\label{sec:arriveGas}

Finally, we consider the scale parameter to be time-varying. In the \emph{generalized autoregressive score (GAS)} framework of \cite{Creal2013}, the time-varying parameters are linearly dependent on their lagged values and the lagged values of the score of the conditional density. Typically, only the first lag is utilized. In our case, the parameter $\alpha_i$ follows recursion
\begin{equation}
\label{eq:gasRecursion}
\begin{aligned}
\alpha_{i+1} &= c + b \alpha_i + a \nabla_{\alpha}(y_i, \alpha_i, \psi, \varphi) \\
&= c + b \alpha_i + a \varphi \left( y_i^{\varphi} e^{-\varphi \alpha_i} - \psi \right), \\
\end{aligned}
\end{equation}
where $c$ is the constant parameter, $b$ is the autoregressive parameter, $a$ is the score parameter and $\nabla_{\alpha}(y_i, \alpha_i, \psi, \varphi)$ is the score defined in \eqref{eq:gengammaScore}. In the GAS framework, the score can be scaled by the inverse of the Fisher information or the square of the inverse of the Fisher information. In our case, however, both scaling functions based on the Fisher information and the unit scaling as well lead to the same model as the Fisher information does not depend on $\alpha_i$. The score for time-varying parameter $\alpha_i$ is the gradient of the log-likelihood with respect to $\alpha_i$ and indicates how sensitive the log-likelihood is to parameter $\alpha_i$. In the GAS model, the score drives the time variation in parameter $\alpha_i$ based on the shape of the generalized gamma density function.

Let $\theta = ( c, b, a, \psi, \varphi)$ denote the vector of parameters in the model. We can estimate $\theta$ straightforwardly by the maximum likelihood method. The log-likelihood function is given by
\begin{equation}
\label{eq:lik}
\ell ( \theta ) = \ln f(y_0 | \alpha_0, \psi, \varphi) + \sum_{i=1}^n \ln f(y_i | \alpha_i, \psi, \varphi),
\end{equation}
where $f(\cdot)$ is the generalized gamma density function given by \eqref{eq:gengammaDensity}. We deliberately set aside the first term as the time-varying parameter $\alpha_i$ needs to be initialized at $i = 0$. We set the value of $\alpha_0$ to the long-term mean value $c / (1 - b)$. Subsequent values of $\alpha_i$, $i = 1, \ldots,n$, than follow recursion \eqref{eq:gasRecursion}. Finally, the parameter estimates $\hat{\theta}$ are obtained by non-linear optimization problem
\begin{equation}
\label{eq:mle}
\hat{\theta} \in \max_{\theta} \ell ( \theta ).
\end{equation}

\section{Empirical Evidence}
\label{sec:emp}

\subsection{Data Overview and Preparation}
\label{sec:empData}

The data sample is obtained from the database of an online bookshop with one brick-and-mortar location in Prague, Czechia. The data cover the period of June 8 -- December 20, 2018, resulting in \num{28} full weeks and \num{5753} observations. The precision of the timestamp is one minute. Thus, zero inter-arrival times might occur in the data due to two or more orders that arrive within one minute. Since the generalized gamma distribution has strictly positive support, the zero inter-arrival times are set to a small positive number. \cite{Bauwens2006} replaces the zero inter-arrival times with a value equal to the half of the minimal positive inter-arrival time and argued that this is a more correct approach than their discarding. Hence, all 81 zero inter-arrival times are set to 0.5 minutes accordingly.

\subsection{Diurnal and Seasonal Patterns}
\label{sec:empSeason}

\begin{figure}
\begin{center}
\includegraphics[width=0.9\textwidth]{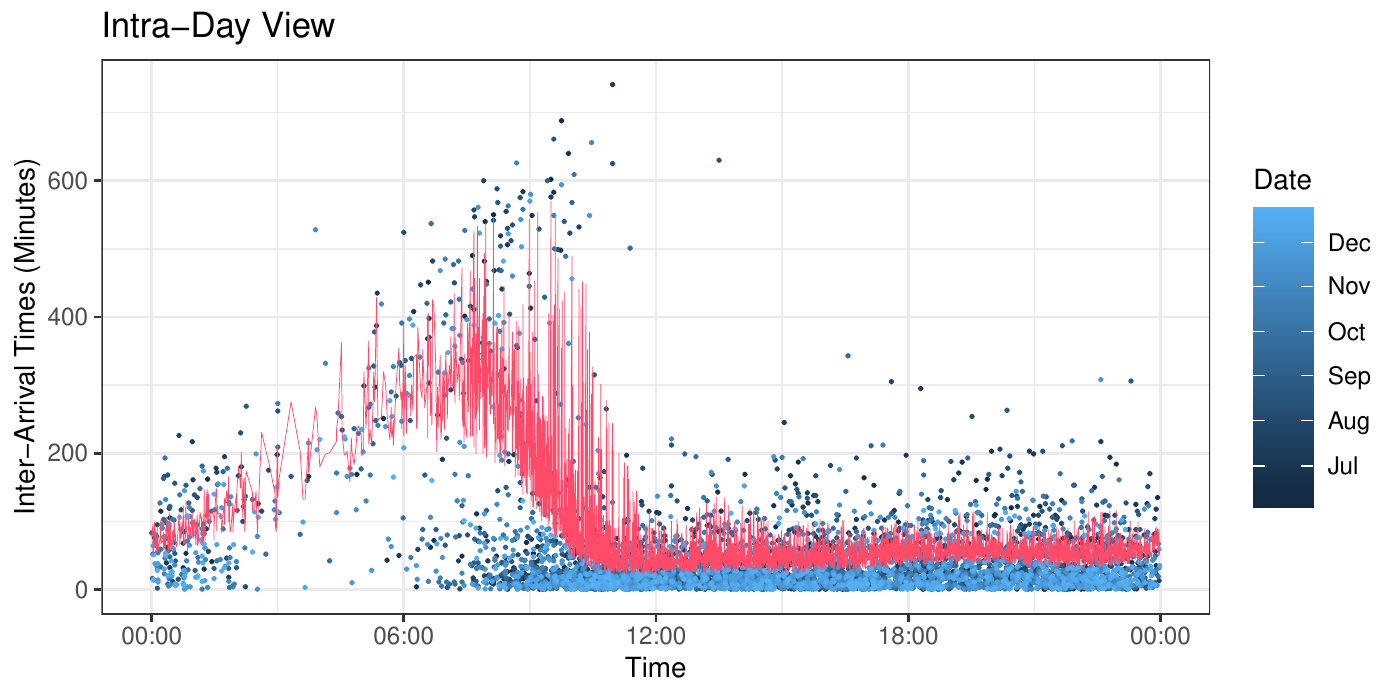}
\caption{Intra-day view of raw inter-arrival times and their fitted diurnal/seasonal pattern.}
\label{fig:durAdjDay}
\end{center}
\end{figure}

\begin{figure}
\begin{center}
\includegraphics[width=0.9\textwidth]{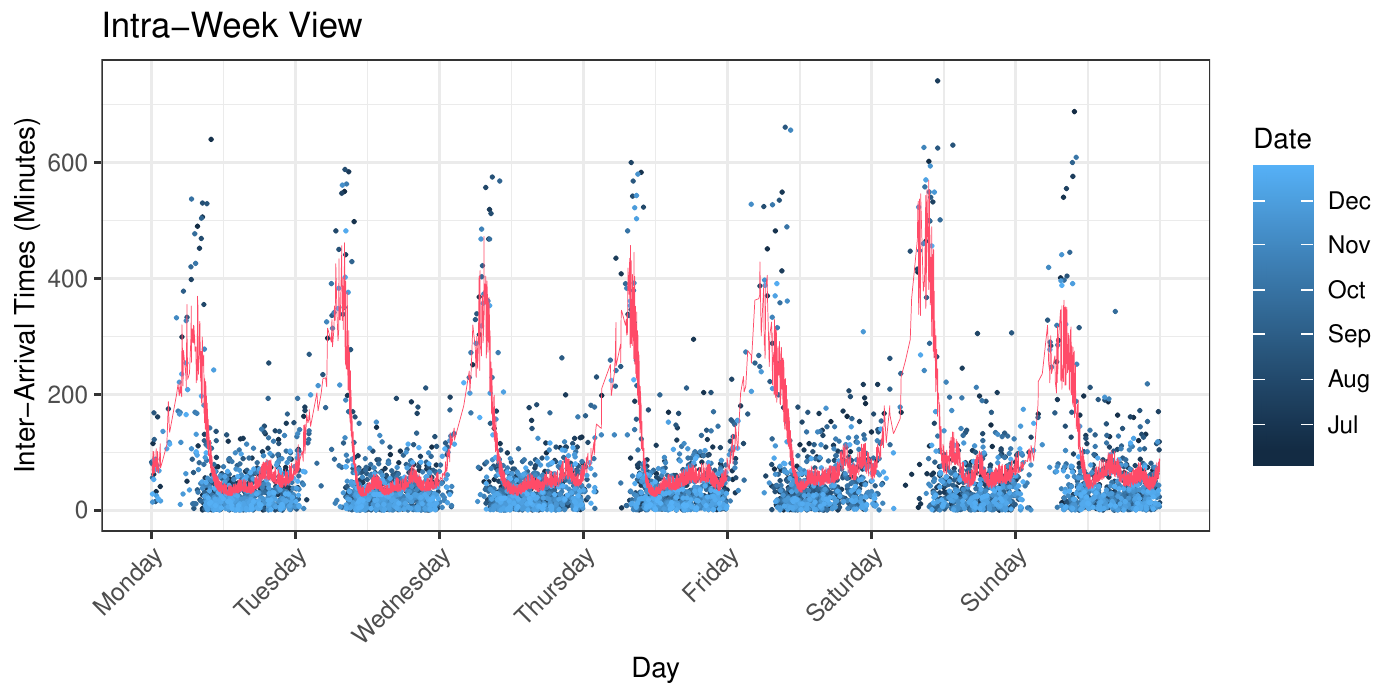}
\caption{Intra-week view of raw inter-arrival times and their fitted diurnal/seasonal pattern.}
\label{fig:durAdjWeek}
\end{center}
\end{figure}

The raw inter-arrival time median is 24 minutes and the mean is 49 minutes -- more than double due to long inter-arrival times during nights (specifically hours between midnight and 9 AM, see Figure \ref{fig:durAdjDay}). Hours between 9 and 11 AM exhibit many short inter-arrival times and several very long inter-arrival times resulting in high dispersion (SD = 111.39). The rest of the rush hours (until 5 PM) shows a similar inter-arrival time median but much lower dispersion (SD = 35.98). Moreover, strong weekly and monthly seasonal patterns are observed. The highest order counts (and consequently lower inter-arrival time values) occur at the beginning of a week and decrease until Saturday, see Figure \ref{fig:durAdjWeek}). On Sundays, order counts increase again and exhibit the highest dispersion. During the summer months, the order counts are rather low (resulting in higher inter-arrival time values) and linearly increase until December.

To obtain the diurnally and seasonally adjusted and detrended inter-arrival times, the regression equation \eqref{eq:splines} with a selected number of knots is estimated. In practice, the selection of a suitable number of knots is an empirically-driven task. Bearing in mind, that too large number of knots can result in overfitting (e.g. one knot for every hour results in too unnatural \emph{bumpy} behavior), on the other hand, that too low number of knots can result in insufficient fit (e.g. one knot for every 2 hours does not capture the nonlinear behavior of data satisfactorily), and after a little experimenting, we select one knot for every 90 minutes which captures all important nonlinearities and does not indicate overfitting. Note that the weekly aggregation is utilized in \eqref{eq:splines} which results in the same daily seasonal component for Mondays, Tuesdays, etc. To ensure continuity between Sundays and Mondays, the sample is stacked three times consecutively and the adjusted inter-arrival times are computed based on the second sub-sample. Parameters are estimated by the WLS.

The fitted values are shown in Figure \ref{fig:durAdjDay} and \ref{fig:durAdjWeek}. Note that they do not coincide with the smooth cubic spline function due to a linear trend which makes the corresponding fitted line \emph{"saw-toothed"}. The diurnally and seasonally adjusted and detrended inter-arrival times are computed as the exponentiated residuals of estimated equation \eqref{eq:splines} and for convenience, they are standardized to have unit mean. Their values range from 0.002 to 11.23 minutes.

\subsection{Fit of the Dynamic Model}
\label{sec:empFit}

\begin{figure}
\begin{center}
\includegraphics[width=0.9\textwidth]{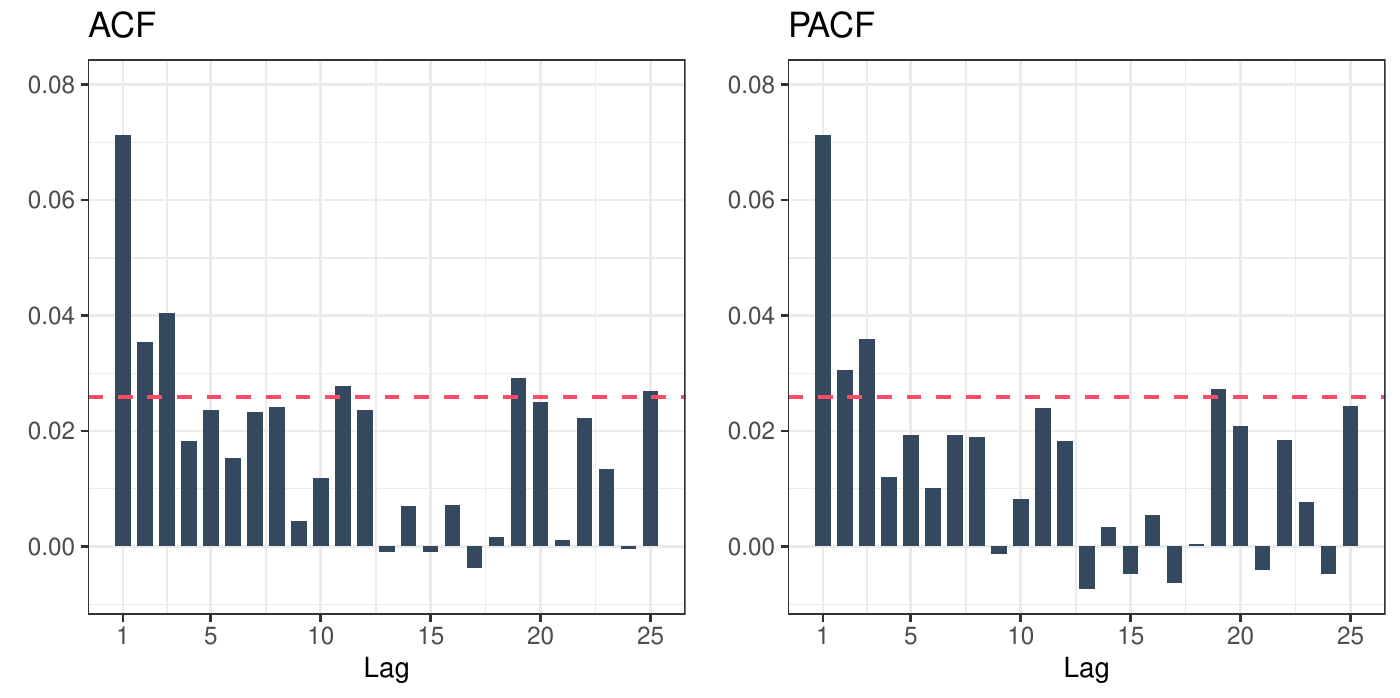}
\caption{The autocorrelation function (ACF) and the partial autocorrelation function (PACF) of adjusted inter-arrival times. Red dashed lines indicate 5\% confidence bounds.}
\label{fig:acfPacf}
\end{center}
\end{figure}

Even after the seasonal and diurnal adjustment, the inter-arrival times still tend to cluster over time -- long (short) inter-arrival times are likely to be followed by long (short) inter-arrival times. This dependence is not particularly strong but nevertheless, it is statistically significant as illustrated in Figure \ref{fig:acfPacf}. To capture the autocorrelation, we utilize the dynamic model based on the generalized gamma distribution with the GAS dynamics in \eqref{eq:gasRecursion}. The parameters are estimated by the maximum likelihood method determined by the non-linear optimization problem in \eqref{eq:mle} and the log-likelihood function in \eqref{eq:lik}. For comparison, we also report the results for static and dynamic models based on special cases of the generalized gamma distribution (G.G.), namely for the exponential (Exp.), Weibull and gamma distributions. 

Parameter estimates and the performance evaluation in terms of the \emph{Akaike information criterion (AIC)} of both static and dynamic inter-arrival time models are shown in Table \ref{tab:fit}. The AIC values are at least by 43.59 lower for dynamic models than for their static counterparts. However, the differences among dynamic models are not so striking -- the highest difference is between exponential and generalized gamma distributions (by 5.94). The best performing model is the most general one -- the dynamic GAS model utilizing the generalized gamma distribution. The dynamic models based on either the exponential or generalized gamma distributions in comparison with their static counterparts are further analyzed in the simulation study of queueing systems. 

\begin{table}
\begin{center}
\begin{tabular}{llrrrrrrr}
\toprule
\multicolumn{2}{c}{Model} & \multicolumn{5}{c}{Estimate} & \multicolumn{2}{c}{Model Fit} \\
Spec. & Dist. & $c$ & $b$ & $a$ & $\psi$ & $\varphi$ & Lik. &  AIC \\
\midrule
Static & Exp.     & 0.00  & \textcolor{gray}{0.00} & \textcolor{gray}{0.00} & \textcolor{gray}{1.00} & \textcolor{gray}{1.00} & \num{-5753.00} & \num{11508.00} \\
Static & Weibull & -0.01 & \textcolor{gray}{0.00} & \textcolor{gray}{0.00} & \textcolor{gray}{1.00} & 0.97 & \num{-5748.93} & \num{11501.86} \\
Static & Gamma   & 0.04  & \textcolor{gray}{0.00} & \textcolor{gray}{0.00} & 0.96 & \textcolor{gray}{1.00} & \num{-5749.77} & \num{11503.54} \\
Static & G. G.   & -0.12 & \textcolor{gray}{0.00} & \textcolor{gray}{0.00} & 1.08 & 0.93 & \num{-5748.37} & \num{11502.75} \\
Dyn.   & Exp.     & 0.00  & 0.76 & 0.06 & \textcolor{gray}{1.00} & \textcolor{gray}{1.00} & \num{-5728.28} & \num{11462.56} \\
Dyn.   & Weibull & 0.00  & 0.75 & 0.06 & \textcolor{gray}{1.00} & 0.97 & \num{-5724.89} & \num{11457.79} \\
Dyn.   & Gamma   & 0.01  & 0.76 & 0.06 & 0.97 & \textcolor{gray}{1.00} & \num{-5725.97} & \num{11459.95} \\
Dyn.   & G. G.   & -0.06 & 0.72 & 0.07 & 1.15 & 0.90 & \num{-5723.31} & \num{11456.62} \\ 
\bottomrule
\end{tabular}
\caption{Parameter estimates of the inter-arrival time models with the log-likelihood value (Lik.) and the Akaike information criterion (AIC).}
\label{tab:fit}
\end{center}
\end{table}

\section{Impact on Queueing Systems}
\label{sec:queue}

\subsection{System with Single Server}
\label{sec:queueSingle}

We investigate the effects of various arrival models on performance measures in queueing systems using simulations. We consider models based on the exponential and generalized gamma distributions with the static and dynamic specifications. The coefficients of the models are taken from Table \ref{tab:fit}. In all models, the rate of arrivals is $\lambda = 1$ job per minute. First, we focus on the queueing system with a single server only. We consider the service times to be independent and exponentially distributed with the rate $\mu$ ranging from $1.1$ to $1.5$ jobs per minute. We simulate the arrival and service processes and measure the number of customers in the system, the busy period of the server, and the response time. The number of simulation runs is equal to $10^9$ which seems to be sufficient for the reported precision of one decimal place as the results are in line with the theoretical performance measures for the static exponential scenario as well as the Little's law for all scenarios.

\begin{figure}
\begin{center}
\includegraphics[width=0.9\textwidth]{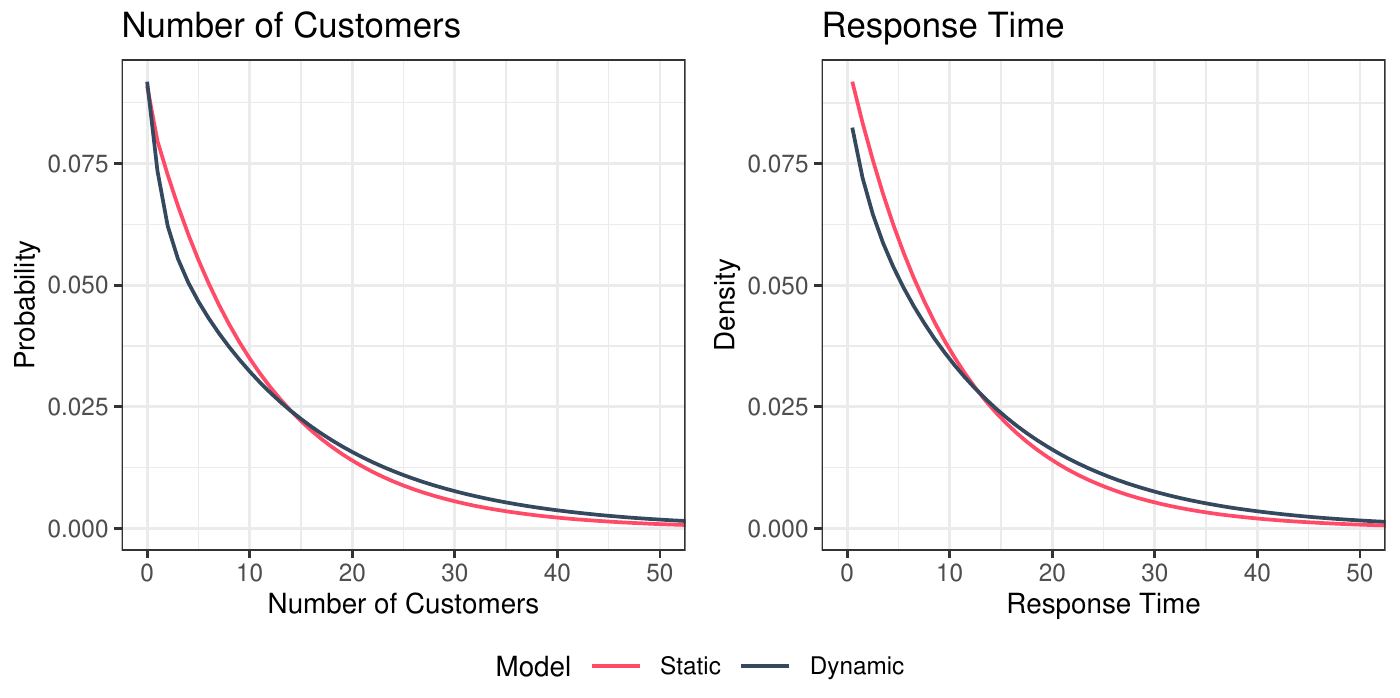}
\caption{The probability mass functions of the number of customers in the system and density functions of the response time for the static and dynamic arrival models based on the generalized gamma distribution in the queueing system with single server and $\mu=1.1$ jobs per minute.}
\label{fig:singleDensity}
\end{center}
\end{figure}

The results are reported in Table \ref{tab:single}. For all values of $\mu$, the systems based on the generalized gamma distribution have higher values of performance measures than the systems based on the exponential distribution in terms of the mean, standard deviation, and $95$ percent quantile. Similarly, systems with the dynamic specification have higher values of performance measures than the systems with the static specification. The left plot of Figure \ref{fig:singleDensity} shows how the probability mass function of the number of customers differs for the static and dynamic models. The dynamic model has a higher probability of the empty system as there tend to be longer periods of low activity. It has also higher probabilities of large numbers of customers in the system as arrivals tend to cluster. The right plot of Figure \ref{fig:singleDensity} shows how the density function of the response time differs for the static and dynamic models. In the dynamic model, customers simply have to wait longer. The differences between the static and dynamic models are naturally weaker for larger $\mu$.

These results carry a warning for practice. When the standard M/M/1 system is assumed but the arrivals actually follow the GAS model based on the generalized gamma distribution, the performance measures are significantly underestimated. For example, the mean number of customers and the mean response time are $22$ percent lower than the actual value for $\mu = 1.1$ jobs per minute. It is therefore crucial to correctly specify the model for arrivals.

\begin{table}
\begin{center}
\begin{tabular}{lllrrrrrrrrr}
\toprule
\multicolumn{3}{c}{Queueing System} & \multicolumn{3}{c}{No. of Customers} & \multicolumn{3}{c}{Busy Period} & \multicolumn{3}{c}{Response Time} \\
$\mu$ & Spec. & Dist. & M & SD & 95\% & M & SD & 95\% & M & SD & 95\% \\
\midrule
  1.1 & Static & Exp. & 10.0 & 10.5 & 31.0 & 10.0 & 45.8 & 39.8 & 10.0 & 10.0 & 30.0 \\ 
  1.1 & Static & G. G. & 10.4 & 10.9 & 32.0 & 10.4 & 47.6 & 41.4 & 10.4 & 10.4 & 31.1 \\ 
  1.1 & Dyn. & Exp. & 12.4 & 13.4 & 39.0 & 10.8 & 54.4 & 41.2 & 12.4 & 12.6 & 37.6 \\ 
  1.1 & Dyn. & G. G. & 12.8 & 13.8 & 41.0 & 11.2 & 56.1 & 43.0 & 12.8 & 13.1 & 39.0 \\ 
  1.2 & Static & Exp. & 5.0 & 5.5 & 16.0 & 5.0 & 16.6 & 22.1 & 5.0 & 5.0 & 15.0 \\ 
  1.2 & Static & G. G. & 5.2 & 5.7 & 17.0 & 5.2 & 17.2 & 22.9 & 5.2 & 5.2 & 15.5 \\ 
  1.2 & Dyn. & Exp. & 6.0 & 6.8 & 20.0 & 5.4 & 19.5 & 23.3 & 6.0 & 6.1 & 18.3 \\ 
  1.2 & Dyn. & G. G. & 6.2 & 7.1 & 20.0 & 5.6 & 20.2 & 24.4 & 6.2 & 6.4 & 19.0 \\ 
  1.3 & Static & Exp. & 3.3 & 3.8 & 11.0 & 3.3 & 9.2 & 14.8 & 3.3 & 3.3 & 10.0 \\ 
  1.3 & Static & G. G. & 3.4 & 3.9 & 11.0 & 3.4 & 9.6 & 15.3 & 3.4 & 3.4 & 10.3 \\ 
  1.3 & Dyn. & Exp. & 3.9 & 4.6 & 13.0 & 3.5 & 10.7 & 15.7 & 3.9 & 4.0 & 11.9 \\ 
  1.3 & Dyn. & G. G. & 4.0 & 4.8 & 14.0 & 3.7 & 11.2 & 16.4 & 4.0 & 4.2 & 12.4 \\ 
  1.4 & Static & Exp. & 2.5 & 3.0 & 8.0 & 2.5 & 6.1 & 11.0 & 2.5 & 2.5 & 7.5 \\ 
  1.4 & Static & G. G. & 2.6 & 3.1 & 9.0 & 2.6 & 6.3 & 11.3 & 2.6 & 2.6 & 7.7 \\ 
  1.4 & Dyn. & Exp. & 2.8 & 3.5 & 10.0 & 2.6 & 7.0 & 11.5 & 2.8 & 2.9 & 8.7 \\ 
  1.4 & Dyn. & G. G. & 3.0 & 3.7 & 10.0 & 2.7 & 7.3 & 12.1 & 3.0 & 3.1 & 9.1 \\ 
  1.5 & Static & Exp. & 2.0 & 2.4 & 7.0 & 2.0 & 4.5 & 8.6 & 2.0 & 2.0 & 6.0 \\ 
  1.5 & Static & G. G. & 2.1 & 2.5 & 7.0 & 2.1 & 4.6 & 8.9 & 2.1 & 2.1 & 6.2 \\ 
  1.5 & Dyn. & Exp. & 2.2 & 2.9 & 8.0 & 2.1 & 5.1 & 9.0 & 2.2 & 2.3 & 6.8 \\ 
  1.5 & Dyn. & G. G. & 2.3 & 3.0 & 8.0 & 2.2 & 5.3 & 9.4 & 2.3 & 2.4 & 7.1 \\ 
 \bottomrule
\end{tabular}
\caption{Mean values (M), standard deviations (SD) and 95\%-quantiles (95\%) of the number of customers in the system, the busy period of the server and the response time in various queueing systems with a single server.}
\label{tab:single}
\end{center}
\end{table}

\subsection{System with Multiple Servers}
\label{sec:queueMulti}

Next, we consider queueing systems with multiple servers. We base the simulations on the same setting as in the previous section. The only difference lies in the service structure. We consider the number of servers $c$ ranging from $11$ to $15$ with the individual service rate $\mu = 0.1$ jobs per minute. Such values result in the same server utilizations $\rho = \lambda / (c \mu)$ as in the previous section. Again, we measure the number of customers in the system, the busy period of the servers, and the response time. By the busy period, we mean the full busy period, i.e. the duration of the state in which all servers are busy.

The results are reported in Table \ref{tab:multi}. The findings are very similar to the system with a single server -- the generalized gamma distribution and the dynamic specification increase all performance measures. When incorrectly assuming the M/M/c system, the specification error is distinct but not as high as in the case of a single server. For example, when assuming the M/M/11 system, the mean number of customers and the mean response time are $14$ percent lower than the actual value for arrivals based on the generalized gamma distribution with the dynamic specification.

\begin{figure}
\begin{center}
\includegraphics[width=0.9\textwidth]{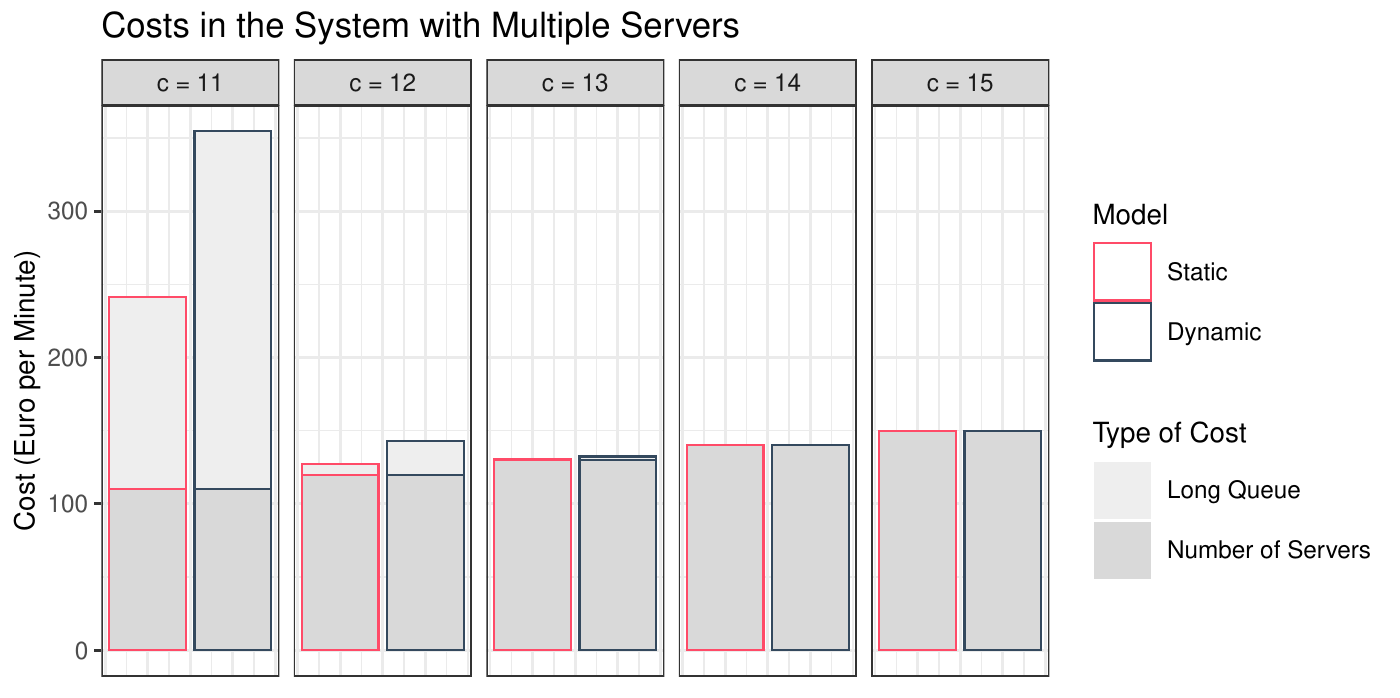}
\caption{Costs related to the number of servers and long queue for the static and dynamic arrival models based on the generalized gamma distribution in the queueing systems with multiple servers and $\mu=0.10$ jobs per minute.}
\label{fig:multiCosts}
\end{center}
\end{figure}

We illustrate how the misspecification of the arrival model can affect decision making in the following toy example. Let us consider that there are two types of costs associated with the operation of the system -- the cost of running one server per unit of time $C_1 = 10$ euro per minute and the cost of the queue longer than $30$ customers per unit of time $C_2 = \num{3000}$ euro per minute. The analytic department is faced with the question of how many servers to operate. The composition of costs for different numbers of servers is shown in Figure \ref{fig:multiCosts}. The optimal number of servers according to the static model is $12$ while it is $13$ for the dynamic model. An analyst assuming the static model believes that the total optimal costs are $127.13$ euro per minute while they actually are $142.87$ euro per minute for the suboptimal choice of 12 servers. An analyst correctly specifying the dynamic model finds out that the lowest possible costs are $132.32$ euro per minute for the optimal choice of 13 servers. The decision based on the misspecified arrival model therefore results in a total cost increase of $8$ percent.

\begin{table}
\begin{center}
\begin{tabular}{lllrrrrrrrrr}
\toprule
\multicolumn{3}{c}{Queueing System} & \multicolumn{3}{c}{No. of Customers} & \multicolumn{3}{c}{Busy Period} & \multicolumn{3}{c}{Response Time} \\
$c$ & Spec. & Dist. & M & SD & 95\% & M & SD & 95\% & M & SD & 95\% \\
\midrule
  11 & Static & Exp. & 16.8 & 10.7 & 38.0 & 10.0 & 45.8 & 39.9 & 16.8 & 13.8 & 43.8 \\ 
  11 & Static & G. G. & 17.2 & 11.1 & 39.0 & 10.4 & 47.6 & 41.4 & 17.2 & 14.0 & 44.6 \\ 
  11 & Dyn. & Exp. & 19.1 & 13.5 & 46.0 & 12.4 & 58.6 & 49.7 & 19.1 & 15.7 & 49.9 \\ 
  11 & Dyn. & G. G. & 19.5 & 14.0 & 47.0 & 12.8 & 60.3 & 51.8 & 19.5 & 16.1 & 50.9 \\ 
  12 & Static & Exp. & 12.2 & 5.8 & 24.0 & 5.0 & 16.6 & 22.1 & 12.2 & 10.8 & 33.6 \\ 
  12 & Static & G. G. & 12.4 & 6.1 & 24.0 & 5.2 & 17.2 & 22.9 & 12.4 & 10.9 & 33.8 \\ 
  12 & Dyn. & Exp. & 13.1 & 7.2 & 27.0 & 6.1 & 21.1 & 27.5 & 13.1 & 11.3 & 35.4 \\ 
  12 & Dyn. & G. G. & 13.3 & 7.5 & 28.0 & 6.3 & 21.9 & 28.7 & 13.3 & 11.5 & 35.8 \\ 
  13 & Static & Exp. & 11.0 & 4.4 & 19.0 & 3.3 & 9.2 & 14.8 & 11.0 & 10.3 & 31.3 \\ 
  13 & Static & G. G. & 11.0 & 4.5 & 19.0 & 3.4 & 9.6 & 15.3 & 11.0 & 10.3 & 31.4 \\ 
  13 & Dyn. & Exp. & 11.4 & 5.2 & 21.0 & 4.0 & 11.7 & 18.4 & 11.4 & 10.5 & 32.0 \\ 
  13 & Dyn. & G. G. & 11.5 & 5.4 & 22.0 & 4.2 & 12.1 & 19.1 & 11.5 & 10.5 & 32.2 \\ 
  14 & Static & Exp. & 10.4 & 3.8 & 17.0 & 2.5 & 6.1 & 11.0 & 10.4 & 10.1 & 30.5 \\ 
  14 & Static & G. G. & 10.5 & 3.9 & 17.0 & 2.6 & 6.3 & 11.3 & 10.5 & 10.1 & 30.6 \\ 
  14 & Dyn. & Exp. & 10.7 & 4.4 & 19.0 & 3.0 & 7.7 & 13.5 & 10.7 & 10.2 & 30.9 \\ 
  14 & Dyn. & G. G. & 10.7 & 4.5 & 19.0 & 3.1 & 8.0 & 14.0 & 10.7 & 10.2 & 30.9 \\ 
  15 & Static & Exp. & 10.2 & 3.5 & 16.0 & 2.0 & 4.5 & 8.6 & 10.2 & 10.0 & 30.2 \\ 
  15 & Static & G. G. & 10.2 & 3.6 & 16.0 & 2.1 & 4.6 & 8.9 & 10.2 & 10.0 & 30.2 \\ 
  15 & Dyn. & Exp. & 10.3 & 3.9 & 17.0 & 2.3 & 5.6 & 10.5 & 10.3 & 10.1 & 30.4 \\ 
  15 & Dyn. & G. G. & 10.3 & 4.1 & 18.0 & 2.4 & 5.8 & 10.9 & 10.3 & 10.1 & 30.4 \\ 
\bottomrule
\end{tabular}
\caption{Mean values (M), standard deviations (SD) and 95\%-quantiles (95\%) of the number of customers in the system, the full busy period of servers and the response time in various queueing systems with multiple servers and $\mu=0.1$ jobs per minute.}
\label{tab:multi}
\end{center}
\end{table}

\subsection{Discussion of More Complex Systems}
\label{sec:queueDisc}

In this paper, we focus on rather simple queueing systems in order to get transparent results. The M/M/1 system is as straightforward as it can be and therefore the best choice for an illustration of the impact of autocorrelated arrivals. The M/M/c system is used as a robustness check to show that the behavior observed for the M/M/1 system is present even for different specifications. As for the toy example of decision making in the M/M/c system, it is meant just as a simplistic illustration revealing a potential source of suboptimal decisions.

On the other hand, \cite{Tomanova2018}, \cite{Tomanova2019}, and \cite{Tomanova2019a} explore a much more realistic and complex queueing system specific to our online bookshop case. As this queueing system is tailored just for this specific application and cannot be easily transferred to others, we only summarize the main findings. \cite{Tomanova2018} performs a process quality assessment based on process simulation and reports that the key quality target is not satisfied in almost twice more cases when the dynamic model is considered (the target is not satisfied in 6.16 percent) than when the static model is considered (for which the target is not satisfied in 3.23 percent). The common approach -- a static model which assumes that times between arrivals follow the exponential distribution with a constant rate -- underestimates the probability of extreme values and thus significantly skews the basis for process quality assessment and leads to suboptimal decisions. \cite{Tomanova2019} also demonstrates that the clustering of arrivals increases the probability of weeks with an extreme number of arrivals that has a negative impact on target fulfillment. \cite{Tomanova2019a} further extends the work for final recommendations for the management of the online bookshop. The main finding is that 21 percent of orders are not satisfied within a working day due to insufficiently allocated resources for the first stage (pre-processing of arrivals).

\section{Conclusion}
\label{sec:con}

We analyze the dependence of inter-arrival times in queueing systems and demonstrate the negative impact of arrival model misspecification on decision making. To capture the autocorrelation structure of inter-arrival times, we propose to utilize a dynamic model based on the generalized gamma distribution with the GAS dynamics. We argue that this approach is superior to the standard model assuming the exponential distribution with a constant rate since it leads to a more faithful representation of the mean and extreme values of the arrival process. Our approach consists of three steps.
\begin{enumerate}
\item We construct a suitable model for capturing the diurnal and seasonal dependencies which takes into account a specific time-structure of inter-arrival times. We utilize a cubic spline approach and propose to estimate the parameters by the weighted ordinary least square method to properly adjust inter-arrival times during hours that exhibit a small median but a huge dispersion. 
\item We argue that the GAS models based on the generalized gamma distribution and its special cases fit the data better than their static counterparts. This is due to the fact that the static models ignore the autocorrelation structure which is still present even after the proper diurnal and seasonal adjustment. 
\item We compare both static and dynamic models in the simulation study of queueing systems with single and multiple servers and exponential services. We show that ignoring the autocorrelation structure leads to biased performance measures. The number of customers in the system, the busy period of servers and the response time have higher mean and variance as well as heavier tails for the proposed dynamic arrivals model than for the standard static model. We also demonstrate how the trust in the standard static model for inter-arrival times leads to suboptimal decisions and consequently to a profit loss.
\end{enumerate}

A proper treatment of arrival dependence is of a great importance since its ignorance generates extra costs. Our approach is useful for process simulations and consequently for process optimization and process quality assessment.

The main limitation of the paper and a topic for future research is the theoretical treatment of the queueing systems with inter-arrival times following the GAS model. In the paper, we resort to simulations to determine the moments, quantiles, and density functions of the performance measures. Theoretical derivation of these quantities and functions is undoubtedly challenging but perhaps possible in some cases. Another topic for future research, that is easier to achieve, is the use of the proposed approach in other applications. Besides retail order processing, these may include customer service, project management, manufacturing engineering, emergency services, logistics, transportation, telecommunication, computing, and others.

\section*{Acknowledgements}
\label{sec:acknow}

We would like to thank the organizers and participants of the 7th International Conference on Management (Nový Smokovec, September 26--29, 2018), the 30th European Conference on Operational Research (Dublin, June 23--26, 2019), the 15th International Symposium on Operations Research in Slovenia (Bled, September 25--27, 2019) and the 3rd International Conference on Advances in Business and Law (Dubai, November 23--24, 2019) for fruitful discussions.

\section*{Funding}
\label{sec:fund}

The work on this paper was supported by the Internal Grant Agency of the University of Economics, Prague under project F4/27/2020 and the Czech Science Foundation under project 19-08985S.


\end{document}